\documentclass[preprint,showpacs,showkeys,preprintnumbers,amsmath,amssymb]{revtex4}
 \usepackage{dcolumn}
 \usepackage{bm}

\begin{document}

 \title{ Hawking Temperature of Dilaton Black Holes from Tunneling }

 \author{ Ji-Rong Ren }

 \author{ Ran Li }

 \thanks{Corresponding author. Electronic mail: liran05@lzu.cn}

 \author{ Fei-Hu Liu }

 \affiliation{Institute of Theoretical Physics, Lanzhou University, Lanzhou, 730000, Gansu, China}

 \begin{abstract}

 Recently, it has been suggested that Hawking radiation can be
 derived from quantum tunnelling methods. In this letter,
 we calculated Hawking temperature of
 dilatonic black holes from tunnelling formalism.
 The two semi-classical methods adopted here are:
 the null-geodesic method proposed by Parikh and Wilczek and
 the Hamilton-Jacobi method propsed by Angheben \textit{et al}.
 We apply the two methods to anylysis the Hawking temperature
 of the static spherical symmetric dilatonic black hole,
 the rotating Kaluza-Klein black hole, and the rotating Kerr-Sen
 black hole.

 \end{abstract}

 \maketitle

 \section{introduction}

 S. W. Hawking\cite{hawking} have been able to derive that black hole can radiate
 from the event horizon like a
 black body at the temperature $T=\frac{\kappa}{2\pi}$ using the
 method of quantum field theory in curved spacetime.
 Rencent proposal of deriving the Hawking temperature via the
 semi-classical methods have been attracted a lot of interests
 \cite{padmanabhan,padmanabhan1,zerbini,zerbini1,arzano,arzano1,arzano2,arzano3,
 parikh,Medved,Medved1,wushuangqing,wushuangqing1,wushuangqing2,wushuangqing3,
 liuwenbiao,zhaozheng,zhaozheng1,zhaozheng2,zhaozheng3,kerner,kerner1,akhmedov,mitra,gui}.
 The complex path method\cite{padmanabhan,padmanabhan1} was first proposed
 by K. Srinivasan and T. Padmanabhan, and subsequently developed
 by many authors \cite{arzano,arzano1,arzano2,arzano3,zerbini,zerbini1}. The developed
 method is called Hamilton-Jacobi method in which the Hawking
 temperature was obtained from calculating the imaginary part of
 classical action toward Hamilton-Jacobi equation. Another method
 called null-geodesic method
 was propsed by M. K. Parikh and F. Wilczek in \cite{parikh}.
 There have been considerable efforts to generalize this method to
 those of various black hole solutions.
 In this letter, we will
 apply the two method mentioned above to analysis the Hawking
 temperature of dilatonic black hole.
 Our prime motivation is to
 understand the applicability of the two methods to dilaton
 black holes in string theory.
 The Hawking radiation of three kinds of black holes investigated in this letter have
 been studied in \cite{jiang} using the gravitational anomaly
 cancellation mechanism firstly proposed in \cite{anomaly}.
 The discussion of thermodynamics of some black holes in dilaton gravity
 can be found in \cite{Astefanesei,Astefanesei1}.
 Let us begin with the spherical
 symmetric dilaton black hole.

 \section{Hawking temperature of spherical
 symmetric dilaton black hole }

 The action for the dilaton gravity describing the dilaton field
 coupled to the $U(1)$ gauge field in $(3+1)$ dimensional is
 subject to the form
 \begin{eqnarray}
 S&=&\frac{1}{16\pi}\int dx^4\sqrt{-g}
 \big[R-2\nabla^\mu\phi\nabla_\mu\phi\nonumber\\
 &&-e^{-2\alpha\phi}F_{\mu\nu}F^{\mu\nu}\big]\;,
 \end{eqnarray}
 where $\phi$ is dilaton field and
 $F_{\mu\nu}=2\nabla_{[\mu}A_{\nu]}$ is the $U(1)$ gauge field
 respectively, with an arbitrary coupling constant $\alpha$. From
 the action, the static spherically symmetric solution of motion
 equation for the underlying theory can be written as\cite{GHS}
 \begin{eqnarray}
 &&ds^2=-\frac{\Delta}{R^2}dt^2+\frac{R^2}{\Delta}dr^2+R^2(d\theta^2+\textrm{sin}^2\theta
 d\varphi^2)\;,\\
 &&\phi=\frac{\alpha}{1+\alpha^2}\textrm{ln}\Big(1-\frac{r_-}{r}\Big)\;,\\
 &&F=\frac{Q}{r^2}dt\wedge dr\;,
 \end{eqnarray}
 where
 \begin{eqnarray}
 \Delta=(r-r_+)(r-r_-)\;,\;\;
 R=r\Big(1-\frac{r_-}{r}\Big)^{\alpha/(1+\alpha^2)}\;.\nonumber
 \end{eqnarray}
 in which the outer and inner horizons are respectively given by
 \begin{equation}
 r_{\pm}=\frac{1+\alpha^2}{1\pm\alpha^2}\big[M\pm\sqrt{M^2-(1-\alpha^2)Q^2}\big]\;.
 \end{equation}
 After performing the conformal transformation
 $\widetilde{g}_{\mu\nu}=e^{-2\alpha\phi}g_{\mu\nu}$, the line
 element becomes the general spherical symmetric form
 \begin{equation}
 ds^2=-f(r)dt^2+\frac{1}{g(r)}dr^2+r^2(d\theta^2+\textrm{sin}^2\theta
 d\varphi^2)\;,
 \end{equation}
 where
 \begin{equation}
 f(r)=\frac{\Delta}{R^2}e^{-2\alpha\phi}\;,\;\;g(r)=\frac{\Delta}{R^2}e^{2\alpha\phi}\;.
 \end{equation}
 Now, we focus on investigating Hawking temperature
 of the dilatonic black hole from quantum tunnelling process.
 It should be noted that the conformal transformation does not
 affect the final result as shown in the following analysis.
 To apply the null-geodesics method,
 it is necessary to choose coordinates which
 are not singular at the horizon.
 This new coordinate has been
 systematically studied by Maulik K. Parikh in \cite{parikhplb}.
 Introduce the coordinate transformation
 \begin{equation}
 dt=dT-\Lambda(r)dr\;,
 \end{equation}
 where the function $\Lambda$ is required to depend only on $r$
 not $t$. Then the line element becomes
 \begin{eqnarray}
 ds^2&=&-f(r)dT^2+\Big(\frac{1}{g(r)}-f(r)\Lambda(r)\Big)dr^2\nonumber\\
 &&+2f(r)\Lambda(r)dTdr+r^2d\Omega^2\;.
 \end{eqnarray}
 Restrict the condition
 \begin{equation}
 \frac{1}{g(r)}-f(r)\Lambda(r)=1\;.
 \end{equation}
 One can obtain the line element in new coordinate
 \begin{eqnarray}
 ds^2&=&-f(r)dT^2+2\sqrt{\frac{f(r)(1-g(r))}{g(r)}}dTdr\nonumber\\
 &&+dr^2+r^2d\Omega^2\;.
 \end{eqnarray}
 This coordinate system has a number of interesting
 features. At any fixed time the spatial geometry is flat. At any
 fixed radius the boundary geometry is the same as that of the
 original metric.

 The radial null geodesics for this metric is given by
 \begin{equation}
 \dot{r}=\sqrt{\frac{f(r)}{g(r)}}\big(\pm 1-\sqrt{1-g(r)}\big)\;.
 \end{equation}
 The imaginary part of the classical action
 for an outgoing positive energy particle is
 \begin{equation}
 \textrm{Im}S=\textrm{Im}\int_{r_{in}}^{r_{out}}p_r dr
 =\textrm{Im}\int_{r_{in}}^{r_{out}}\int_{0}^{p_r}dp_r'dr\;.
 \end{equation}
 where $r_{in}$ and $r_{out}$ are the initial and final radii of
 the black hole, respectively. Assume that the emitted energy $\omega'\ll
 M$. According to energy conservation, the energy of background
 spacetime $M$ becomes $M-\omega'$. From Hamilton equation
 $\dot{r}=\frac{dH}{dp_r}|_r$, the integral can be rewritten as
 \begin{equation}
 \textrm{Im}S=\textrm{Im}\int_{r_{in}}^{r_{out}}\int_{M}^{M-\omega}\frac{dr}{\dot{r}}dH\;,
 \end{equation}
 where $dH=-d\omega'$ because $H=M-\omega'$.
 In order to find the Hawking temperature, we can perform a series
 expansion in $\omega$. The first order gives
 \begin{eqnarray}
 \textrm{Im}S&=&\textrm{Im}\int_{r_{in}}^{r_{out}}\int_{M}^{M-\omega}\frac{dr}{\dot{r}(r,M-\omega')}(-d\omega')\nonumber\\
 &=&-\omega\textrm{Im}\int_{r_{in}}^{r_{out}}\frac{dr}{\dot{r}(r,M)}+O(\omega^2)\nonumber\\
 &\simeq&\omega\textrm{Im}\int_{r_{out}}^{r_{in}}\frac{dr}{\dot{r}(r,M)}\;.
 \end{eqnarray}
 To proceed further we will need to estimate the last integral
 which
 can be done by deforming the contour.
 There is a simple pole at the horizon with a residue
 $\frac{2}{\sqrt{f'(r_+)g'(r_+)}}$.
 Hence the imaginary part of the action will be
 \begin{equation}
 \textrm{Im}S=\frac{2\pi\omega}{\sqrt{f'(r_+)g'(r_+)}}\;.
 \end{equation}
 Using the WKB approximation the tunneling probability
 for the classically forbidden trajectory is given by
 \begin{equation}
 \Gamma=\textrm{exp}(-2\textrm{Im}S)\;.
 \end{equation}
 Hartle and Hawking in \cite{hartle} obtained
 particle production in the standard
 black-hole spacetimes using a semiclassical analysis.
 The tunneling probability can also be written as
 \begin{equation}
 \Gamma=\textrm{exp}(-\beta\omega)\;,
 \end{equation}
 where $\omega$ is the energy of the particles and $\beta^{-1}$ is
 the Hawking temperature.
 The higher order terms are a self-interaction effect. For calculating the
 Hawking temperature, expansion to linear order is all that is required.
 Comparing the above equation,
 the hawking temperature is given by
 \begin{equation}
 T_H=\beta^{-1}=\frac{\sqrt{f'(r_+)g'(r_+)}}{4\pi}\;.
 \end{equation}
 It should be noted that the Hawking temperature is depend on the
 product of $f'(r)$ and $g'(r)$ which is independent of the conformal
 transformation. Due to the same reason, it is shown that the Hawking temperature
 obtained from the Hamilton-Jacobi method is also independent of
 the conformal transformation. The tunneling method we applied to
 the spherical symmetric dilaton black hole is not affected by the
 conformal transformation.
 For the static spherically symmetric dilatonic black hole in our
 case, this gives the hawking temperature
 \begin{equation}
 T_H=\frac{1}{4\pi
 r_+}(1-\frac{r_-}{r_+})^{(1-\alpha^2)/(1+\alpha^2)}\;.
 \end{equation}

 We now turn to the Hamilton-Jacobi method.
 Consider a massive scalar field in the static spherically
 symmetric dilatonic black hole spacetime
 satisfying Klein-Gordon equation
 \begin{equation}
 g^{\mu\nu}\nabla_{\mu}\nabla_{\nu}\Phi-m^2\Phi=0\;.
 \end{equation}
 By performing the WKB approximation, \textit{i.e.}, expanding the
 field function as
 \begin{equation}
 \Phi=\textrm{exp}(-\frac{i}{\hbar}S+\cdots)\;,
 \end{equation}
 one can obtain Hamilton-Jacobi equation
 \begin{equation}
 g^{\mu\nu}\partial_\mu S\partial_\nu S+m^2=0\;.
 \end{equation}
 where $S$ is the classical action. For the metric of the form
 \begin{equation}
 ds^2=-f(r)dt^2+\frac{1}{g(r)}dr^2+h_{ij}dx^idx^j\;,
 \end{equation}
 where $h_{ij}dx^idx^j=r^2d\Omega^2$ in the case of the static spherically
 symmetric dilatonic black hole spacetime.
 The Hamilton-Jacobi equation becomes
 \begin{equation}
 -\frac{(\partial_t S)^2}{f(r)}+g(r)(\partial_r
 S)^2+h^{ij}\partial_i S\partial_j S+m^2=0\;.
 \end{equation}
 One can use separation of variables to
 write the solution of the form
 \begin{equation}
 S=-Et+W(r)+J(x^i)\;.
 \end{equation}
 As a consequence, one get
 \begin{equation}
 \partial_t S=-E\;,\;\;\partial_r
 S=W'(r)\;,\;\;\partial_{i}S=J_i\;,
 \end{equation}
 where the $J_i$s are constants.
 $W(r)$ can be solved
 \begin{equation}
 W(r)=\int\frac{dr}{\sqrt{f(r)g(r)}}\sqrt{E^2-f(r)(m^2+h^{ij}J_iJ_j)}\;.
 \end{equation}
 Following \cite{arzano}, we can select the proper spatial distant
 \begin{equation}
 d\sigma^2=\frac{dr^2}{g(r)}\;,
 \end{equation}
 where we are only concerned with the radial rays as the
 null-geodesic method.
 Performing the near horizon approximation
 \begin{eqnarray}
 &&f(r)=f'(r_+)(r-r_+)+\cdots\;,\nonumber\\
 &&g(r)=g'(r_+)(r-r_+)+\cdots\;.
 \end{eqnarray}
 we find
 \begin{equation}
 \sigma=\int\frac{dr}{\sqrt{g(r)}}\simeq\frac{2\sqrt{r-r_+}}{\sqrt{g'(r_+)}}\;.
 \end{equation}
 The imaginary part of the classical action is
 \begin{eqnarray}
 \textrm{Im}W(\sigma)&=&\textrm{Im}\frac{2}{\sqrt{g'(r_+)f'(r_+)}}\int\frac{d\sigma}{\sigma}\nonumber\\
 &&\times\sqrt{E^2-\frac{\sigma^2}{4}g'(r_+)f'(r_+)(m^2+h^{ij}J_iJ_j)}\nonumber\\
 &=&\frac{2\pi E}{\sqrt{g'(r_+)f'(r_+)}}\;.
 \end{eqnarray}
 So the imaginary part of classical action calculated by the
 Hamilton-Jacobi method is the same as the previous result from
 the null geodesics method. Now, we will turn to the rotating
 dilatonic black hole.

 \section{Hawking temperature of Kaluza-Klein
 dilaton black hole }

 The Kaluza-Klein black hole is an exact solution of the dilatonic
 action with the coupling constant $\alpha=\sqrt{3}$.
 It is derived by a dimensional reduction
 of the boosted five-dimensional Kerr solution to four dimensions.
 The metric is given by\cite{KK,KK1}
 \begin{eqnarray}
 &&ds^2=-f(r,\theta)dt^2+\frac{1}{g(r,\theta)}dr^2-2H(r,\theta)dtd\varphi\nonumber\\
 &&\;\;\;\;\;\;\;\;\;+K(r,\theta)d\varphi^2+\Sigma(r,\theta)d\theta^2\;,\nonumber\\
 &&f(r,\theta)=\frac{\Delta-a^2\textrm{sin}^2\theta}{B\Sigma}\;,\nonumber\\
 &&g(r,\theta)=\frac{\Delta}{B\Sigma}\;,\nonumber\\
 &&H(r,\theta)=a\textrm{sin}^2\theta\frac{Z}{B\sqrt{1-\nu^2}}\;\nonumber\\
 &&K(r,\theta)=B(r^2+a^2)+a^2\textrm{sin}^2\theta\frac{Z}{B}\;,\nonumber\\
 &&\Sigma(r,\theta)=r^2+a^2\textrm{cos}^2\theta\;,
 \end{eqnarray}
 where
 \begin{eqnarray}
 &&\Delta=r^2-2\mu r+a^2\;,\nonumber\\
 &&Z=\frac{2\mu r}{\Sigma}\;,\nonumber\\
 &&B=\sqrt{1+\frac{\nu^2Z}{1-\nu^2}}\;.
 \end{eqnarray}
 The dilaton field and gauge potential are respectively
 \begin{eqnarray}
 &&\phi=-\frac{\sqrt{3}}{2}\textrm{ln}B\;,\nonumber\\
 &&A_t=\frac{\nu Z}{2(1-\nu^2)B^2}\;,\nonumber\\
 &&A_\varphi=-\frac{a\nu
 Z\textrm{sin}^2\theta}{1\sqrt{1-\nu^2}B^2}\;.
 \end{eqnarray}
 The physical mass $M$, the charge $Q$, and the angular momentum
 $J$ are expressed by the boost parameter $\nu$, mass parameter
 $\mu$, and specific angular momentum $a$, as
 \begin{eqnarray}
 &&M=\mu\Big[1+\frac{\nu^2}{2(1-\nu^2)}\Big]\;,\nonumber\\
 &&Q=\frac{\mu\nu}{1-\nu^2}\;,\nonumber\\
 &&J=\frac{\mu a}{\sqrt{1-\nu^2}}\;.
 \end{eqnarray}
 The outer and inner horizons are respectively given by
 \begin{equation}
 r_\pm=\mu\pm\sqrt{\mu^2-a^2}\;.
 \end{equation}
 The metric can be written as
 \begin{eqnarray}
 ds^2&=&-F(r,\theta)dt^2+\frac{1}{g(r,\theta)}dr^2\nonumber\\
 &&+K(r,\theta)\Big(d\varphi-\frac{H(r,\theta)}{K(r,\theta)}dt\Big)^2
 +\Sigma(r,\theta)d\theta^2\;,
 \end{eqnarray}
 where
 \begin{eqnarray}
 F(r,\theta)&=&f(r,\theta)+\frac{H^2(r,\theta)}{K(r,\theta)}\nonumber\\
 &=&\frac{\Delta[(1-\nu^2)\Sigma+2\mu\nu^2 r]}{B[(1-\nu^2)\Sigma\Delta+2\mu r(r^2+a^2)]}\;.
 \end{eqnarray}
 At the horizon, one have
 \begin{equation}
 \frac{H(r_+,\theta)}{K(r_+,\theta)}=\frac{a\sqrt{1-\nu^2}}{r_+^2+a^2}=\Omega_H\;.
 \end{equation}
 Because the metric depends on the angle $\theta$, we will apply
 the method developed in \cite{kerner,kerner1} to continue. We will first
 fix the angle $\theta$, and then show that the final result is
 independent of the angle $\theta$.
 The metric near the horizon for fixed $\theta=\theta_0$ is
 \begin{eqnarray}
 ds^2&=&-F_r(r_+,\theta_0)(r-r_+)dt^2+\frac{dr^2}{g_r(r_+,\theta_0)(r-r_+)}\nonumber\\
 &&+K(r_+,\theta_0)d\chi^2\;,
 \end{eqnarray}
 where $F_r(r,\theta)$ denotes the partial differential of $F(r,\theta)$, $g_r(r,\theta)$
 denotes the partial differential of $g(r,\theta)$, and $d\chi=d\varphi-\Omega_H
 dt$ is new coordinate parameter. This metric is well behaved for
 all $\theta_0$ and is of the same form as the spherical symmetric
 metric (6) in the last section.
 To see this point more clearly, one can use the drag
 coordinate like this
 \begin{equation}
 \frac{d\varphi}{dt}=\Omega_H\;,
 \end{equation}
 which just means $d\chi=0$.
 Then, the metric can further reduce to the form
 \begin{equation}
 ds^2=-F_r(r_+,\theta_0)(r-r_+)dt^2+\frac{dr^2}{g_r(r_+,\theta_0)(r-r_+)}\;.
 \end{equation}
 According to the procedure in the last section,
 one can easily obtain the final result bying considering the
 massive particle escaping from the horizon
 \begin{equation}
 T_H=\frac{\sqrt{F_r(r_+,\theta_0)g_r(r_+,\theta_0)}}{4\pi}\;.
 \end{equation}
 Direct calculation for $F_r(r_+,\theta_0)$ and
 $g_r(r_+,\theta_0)$ gives
 \begin{eqnarray}
 &&F_r(r_+,\theta_0)=\frac{\sqrt{1-\nu^2}\Delta_r(r_+)}{4\mu^2r_+^2}
 \sqrt{\Sigma(r_+,\theta_0)}\nonumber\\
 &&\;\;\;\;\;\;\;\;\;\;\;\;\;\;
 \times\sqrt{(1-\nu^2)\Sigma(r_+,\theta_0)+2\mu\nu^2r_+}\;,\nonumber\\
 &&g_r(r_+,\theta_0)=\frac{\Delta_r(r_+)}{\sqrt{\Sigma(r_+,\theta_0)}}\nonumber\\
 &&\;\;\;\;\;\;\;\;\;\;\;\;\;\;
 \times\frac{1}{\sqrt{(1-\nu^2)\Sigma(r_+,\theta_0)+2\mu\nu^2r_+}}\;.
 \end{eqnarray}
 Although $F_r(r_+,\theta_0)$ and
 $g_r(r_+,\theta_0)$ each depend on $\theta_0$, their product
 gives the Hawking temperature
 \begin{equation}
 T_H=\frac{1}{2\pi}\frac{\sqrt{1-\nu^2}\sqrt{\mu^2-a^2}}{(r_+^2+a^2)}\;.
 \end{equation}
 which is independent of $\theta_0$.

 Now, we turn to the Hamilton-Jacob method to calculate the
 Hawking temperature. According to metric (41), the action can be
 assumed to of the form
 \begin{equation}
 I=-Et+J\varphi+W(r,\theta_0)\;.
 \end{equation}
 In terms of the relation $\chi(r_+)=\varphi-\Omega_H t$, the classical
 action can be written as
 \begin{equation}
 I=-(E-\Omega_H J)t+J\chi+W(r,\theta_0)\;.
 \end{equation}
 It is easy to obtain the imaginary part of action. In fact, the
 similarity between the metric in this section and the metric in
 the last section reminds us that one can just replace $E$ with $(E-\Omega_H
 J)$ to obtain the imaginary part of classical action. The result
 is given by
 \begin{equation}
 \textrm{Im}W(r,\theta_0)=\frac{2\pi(E-\Omega_H
 J)}{\sqrt{F_r(r_+,\theta_0)g_r(r_+,\theta_0)}}\;.
 \end{equation}
 This in turn gives the same temperature
 \begin{equation}
 T_H=\frac{1}{2\pi}\frac{\sqrt{1-\nu^2}\sqrt{\mu^2-a^2}}{(r_+^2+a^2)}\;.
 \end{equation}
 The two methods used in this letter are also valid to
 the rotating Kaluza-Klein black hole. Now, we wll turn to a more
 general case to analysis this validity.

 \section{Hawking temperature of Kerr-Sen dilaton black hole }

 The Kerr-Sen black hole\cite{kerrsen} is a solution to the
 low-energy effective action in heterotic string theory. The
 action is
 \begin{eqnarray}
 S&=&\frac{1}{16\pi}\int d^4x \sqrt{-g}[R
 -2\nabla^\mu\phi\nabla_\mu\phi\nonumber\\
 &&-e^{-2\phi}F_{\mu\nu}F^{\mu\nu}
 -\frac{1}{12}e^{-4\phi}H^2]\;.
 \end{eqnarray}
 where $H$ is the three-form axion field and the coupling constant
 $\alpha=1$.
 Sen adopted the solution generating technique to obtain a new
 solution from the uncharged Kerr solution. The metric is given by
 \begin{eqnarray}
 &&ds^2=-f(r,\theta)dt^2+\frac{1}{g(r,\theta)}dr^2-2H(r,\theta)dtd\varphi\nonumber\\
 &&\;\;\;\;\;\;\;\;+K(r,\theta)d\varphi^2+\Sigma(r,\theta)d\theta^2\;,\nonumber\\
 &&f(r,\theta)=\frac{\Delta-a^2\textrm{sin}^2\theta}{\Sigma}\;,\nonumber\\
 &&g(r,\theta)=\frac{\Delta}{\Sigma}\;,\nonumber\\
 &&H(r,\theta)=\frac{2\mu ra\textrm{cosh}^2\beta\textrm{sin}^2\theta}{\Sigma}\;\nonumber\\
 &&K(r,\theta)=\frac{\Lambda\textrm{sin}^2\theta}{\Sigma}\;,\nonumber\\
 &&\Sigma(r,\theta)=r^2+a^2\textrm{cos}^2\theta+2\mu r\textrm{sinh}^2\beta\;,
 \end{eqnarray}
 where
 \begin{eqnarray}
 &&\Delta=r^2-2\mu r+a^2\;,\nonumber\\
 &&\Lambda=(r^2+a^2)(r^2+a^2\textrm{cos}^2\theta)+2\mu
 ra^2\textrm{sin}^2\theta\nonumber\\
 &&\;\;\;\;\;\;+4\mu r(r^2+a^2)\textrm{sinh}^2\beta+4\mu^2r^2\textrm{sinh}^4\beta\;.
 \end{eqnarray}
 The dilaton field, axion field, and gauge potential are
 respectively given by
 \begin{eqnarray}
 &&\phi=\frac{1}{2}\textrm{ln}\frac{\Sigma}{r^2+a^2\textrm{cos}^2\theta}\;,\nonumber\\
 &&B_{t\varphi}=2a\textrm{sin}^2\theta\frac{\mu
 r\textrm{sinh}^2\beta}{\Sigma}\;,\nonumber\\
 &&A_t=\frac{\mu
 r\textrm{sinh}2\beta}{\sqrt{2}\Sigma}\;,\nonumber\\
 &&A_\varphi=\frac{a\mu
 r\textrm{sinh}2\beta\textrm{sin}^2\theta}{\sqrt{2}\Sigma}\;.
 \end{eqnarray}
 The mass $M$, the charge $Q$, and the angular momentum $J$ are
 given as
 \begin{eqnarray}
 &&M=\frac{\mu}{2}(1+\textrm{cosh}2\beta)\;,\nonumber\\
 &&Q=\frac{\mu}{\sqrt{2}}\textrm{sinh}^22\beta\;,\nonumber\\
 &&J=Ma\;.
 \end{eqnarray}
 The outer and inner horizons is determined as
 \begin{equation}
 r_\pm=\mu\pm\sqrt{\mu^2-a^2}\;.
 \end{equation}
 As discussed in last section, the metric can be written in the
 form (37). Now, the function $F(r,\theta)$ is given by
 \begin{equation}
 F(r,\theta)=\frac{\Delta\Sigma}{\Lambda}\;.
 \end{equation}
 The angular velocity is
 \begin{equation}
 \Omega_H=\frac{H(r_+,\theta)}{K(r,\theta)}=\frac{a}{(r_+^2+a^2)}\frac{1}{\textrm{cosh}^2\beta}\;.
 \end{equation}
 As shown in the last section, the same procedure of applying the
 two methods in this black hole solution will give the same final
 result
 \begin{equation}
 T_H=\frac{\sqrt{F_r(r_+,\theta_0)g_r(r_+,\theta_0)}}{4\pi}\;.
 \end{equation}
 Direct calculation for $F_r(r_+,\theta_0)$ and
 $g_r(r_+,\theta_0)$ gives
 \begin{eqnarray}
 &&F_r(r_+,\theta_0)=\frac{\Delta_r(r_+)(2\mu r_+\textrm{cosh}^2\beta-a^2\textrm{sin}^2\theta_0)}
 {4\mu^2r_+^2\textrm{cosh}^4\beta}\;,\nonumber\\
 &&g_r(r_+,\theta_0)=\frac{\Delta_r(r_+)}{2\mu
 r_+\textrm{cosh}^2\beta-a^2\textrm{sin}^2\theta_0}\;.
 \end{eqnarray}
 Although $F_r(r_+,\theta_0)$ and
 $g_r(r_+,\theta_0)$ each depend on $\theta_0$, their product
 gives the Hawking temperature
 \begin{equation}
 T_H=\frac{1}{2\pi}\frac{\sqrt{\mu^2-a^2}}{(r_+^2+a^2)\textrm{cosh}^2\beta}\;.
 \end{equation}
 which is independent of $\theta_0$.
 In this section, we see that the null geodesics method and the
 Hamilton-Jacobi method are also valid to the rotating
 Kerr-Sen dilaton black hole solution.

 \section{conclusion}

 In this letter, we have succeeded in
 extending the semi-classical methods to calculate
 the Hawking temperature of dilatonic black holes in string theory.
 The results is consistent with the underlying unitary theory.

 \section*{Acknowledgement}

 The author Ran Li thanks Dr. Tao Zhu for helpful discussions.

 \end{document}